\documentclass[12pt]{article}
\usepackage[colorlinks,linkcolor=Blue,citecolor=Blue,bookmarks,bookmarksnumbered]{hyperref}
\usepackage[scaled=0.85]{helvet}
\usepackage{amsmath,amssymb,accents,mathrsfs,XoohmE}
\usepackage{graphicx,color}
\usepackage{booktabs}
\usepackage{multirow}
\usepackage{placeins}

 \pdfoutput=1

\def\rI{{\rm I}}
\def\rJ{{\rm J}}

\def\rL{{\rm L}}

\def\hi{{\hat\imath}}
\def\hj{{\hat\jmath}}
\def\hk{{\hat{k}}}
\def\hl{{\hat\ell}}

\def\be{\begin{equation}}
\def\ee{\end{equation}}
\newcommand{\bea}{\begin{eqnarray}}
\newcommand{\eea}{\end{eqnarray}}
\newcommand{\ena}{\end{eqnarray}}

\def\pp{{\mathchoice
          {
              \kern 1pt%
              \raise 1pt
              \vbox{\hrule width5pt height0.4pt depth0pt
                    \kern -2pt
                    \hbox{\kern 2.3pt
                          \vrule width0.4pt height6pt depth0pt
                          }
                    \kern -2pt
                    \hrule width5pt height0.4pt depth0pt}%
                    \kern 1pt
           }
            {
              \kern 1pt%
              \raise 1pt
              \vbox{\hrule width4.3pt height0.4pt depth0pt
                    \kern -1.8pt
                    \hbox{\kern 1.95pt
                          \vrule width0.4pt height5.4pt depth0pt
                          }
                    \kern -1.8pt
                    \hrule width4.3pt height0.4pt depth0pt}%
                    \kern 1pt
            }
            {
              \kern 0.5pt%
              \raise 1pt
              \vbox{\hrule width4.0pt height0.3pt depth0pt
                    \kern -1.9pt  
                    \hbox{\kern 1.85pt
                          \vrule width0.3pt height5.7pt depth0pt
                          }
                    \kern -1.9pt
                    \hrule width4.0pt height0.3pt depth0pt}%
                    \kern 0.5pt
            }
            {
              \kern 0.5pt%
              \raise 1pt
              \vbox{\hrule width3.6pt height0.3pt depth0pt
                    \kern -1.5pt
                    \hbox{\kern 1.65pt
                          \vrule width0.3pt height4.5pt depth0pt
                          }
                    \kern -1.5pt
                    \hrule width3.6pt height0.3pt depth0pt}%
                    \kern 0.5pt
            }
        }}

\def\mm{{\mathchoice
                       {
                             \kern 1pt
               \raise 1pt    \vbox{\hrule width5pt height0.4pt depth0pt
                                  \kern 2pt
                                  \hrule width5pt height0.4pt depth0pt}
                             \kern 1pt}
                       {
                            \kern 1pt
               \raise 1pt \vbox{\hrule width4.3pt height0.4pt depth0pt
                                  \kern 1.8pt
                                  \hrule width4.3pt height0.4pt depth0pt}
                             \kern 1pt}
                       {
                            \kern 0.5pt
               \raise 1pt
                            \vbox{\hrule width4.0pt height0.3pt depth0pt
                                  \kern 1.9pt
                                  \hrule width4.0pt height0.3pt depth0pt}
                            \kern 1pt}
                       {
                           \kern 0.5pt
             \raise 1pt  \vbox{\hrule width3.6pt height0.3pt depth0pt
                                  \kern 1.5pt
                                  \hrule width3.6pt height0.3pt depth0pt}
                           \kern 0.5pt}
                       }}

\def\ad{{\kern0.5pt
                   \alpha \kern-5.05pt \raise5.8pt\hbox{$\textstyle.$}\kern
0.5pt}}

\def\bd{{\kern0.5pt
                   \beta \kern-5.05pt \raise5.8pt\hbox{$\textstyle.$}\kern
0.5pt}}

\def\qd{{\kern0.5pt
                   q \kern-5.05pt \raise5.8pt\hbox{$\textstyle.$}\kern
0.5pt}}
\def\Dot#1{{\kern0.5pt
     {#1} \kern-5.05pt \raise5.8pt\hbox{$\textstyle.$}\kern
0.5pt}}

\catcode`@=11
\def\un#1{\relax\ifmmode\@@underline#1\else
        $\@@underline{\hbox{#1}}$\relax\fi}
\catcode`@=12

\def\a{\alpha}
\def\b{\beta}
\def\c{\chi}
\def\d{\delta}

\def\g{\gamma}

\def\i{\iota}
\def\j{\psi}
\def\k{\kappa}
\def\l{\lambda}
\def\m{\mu}
\def\n{\nu}
\def\o{\omega}

\def\r{\rho}
\def\s{\sigma}
\def\t{\tau}

\def\L{\Lambda}
\def\O{\Omega}
\def\P{\Pi}

\def\S{\Sigma}

\def\ca{{\cal A}}

\def\cf{{\cal F}}

\def\dslash{\not{\hbox{\kern-2pt $\partial$}}}
\def\Dslash{\not{\hbox{\kern-4pt $D$}}}
\def\pslash{\not{\hbox{\kern-2.3pt $p$}}}
 \newtoks\slashfraction
 \slashfraction={.13}
 \def\slash#1{\setbox0\hbox{$ #1 $}
 \setbox0\hbox to \the\slashfraction\wd0{\hss \box0}/\box0 }
 
\def\kcr{{\hbox{\ro \char'170}}}                
\def\ktl{{\hbox{\ro \char'170}}}        
\def\ktr{{\hbox{\ro \char'170}}}        
\def\kbl{{\hbox{\ro \char'170}}}        
\def\kbr{{\hbox{\ro \char'170}}}        

\def\plpl{\raise-2pt\hbox{$\raise3pt\hbox{$_+$}\hskip-6.67pt\raise0.0pt
\hbox{$^+$}\hskip 0.01pt$}}
\def\mimi{\raise-2pt\hbox{$\raise3pt\hbox{$_-$}\hskip-6.67pt\raise0.0pt
\hbox{$^-$}\hskip 0.01pt$}} 

\def\bo{{\raise.15ex\hbox{\large$\Box$}}}               
\def\TH{{\raise.2ex\hbox{$\displaystyle \bigodot$}\mskip-4.7mu \llap H \;}}
\def\face{{\raise.2ex\hbox{$\displaystyle \bigodot$}\mskip-2.2mu \llap {$\ddot
        \smile$}}}                                      

\def\dt#1{\on{\hbox{\bf .}}{#1}}                
\def\Dot#1{\dt{#1}}

\def\Tilde#1{\widetilde{#1}}                    
\def\Hat#1{\widehat{#1}}                        
\def\leftrightarrowfill{$\mathsurround=0pt \mathord\leftarrow \mkern-6mu
        \cleaders\hbox{$\mkern-2mu \mathord- \mkern-2mu$}\hfill
        \mkern-6mu \mathord\rightarrow$}
\def\dvec#1{\vbox{\ialign{##\crcr
        \leftrightarrowfill\crcr\noalign{\kern-1pt\nointerlineskip}
        $\hfil\displaystyle{#1}\hfil$\crcr}}}           
\def\dt#1{{\buildrel {\hbox{\LARGE .}} \over {#1}}}     

\def\sfrac#1#2{{\vphantom1\smash{\lower.5ex\hbox{\small$#1$}}\over
        \vphantom1\smash{\raise.4ex\hbox{\small$#2$}}}} 
\def\bfrac#1#2{{\vphantom1\smash{\lower.5ex\hbox{$#1$}}\over
        \vphantom1\smash{\raise.3ex\hbox{$#2$}}}}       
\def\afrac#1#2{{\vphantom1\smash{\lower.5ex\hbox{$#1$}}\over#2}}    

\newcommand{\bm}[1]{{\boldsymbol{#1}}}

\def\ad{{\dot{\alpha}}}
\def\bd{{\dot{\beta}}}

 \font\rOpe=cmsy10                        
 \def\ktl{{\hbox{\rOpe\char'170}}}        
 \def\kbl{{\hbox{\rOpe\char'170}}}        
 \def\kcr{{\reflectbox{\rOpe\char'170}}}        
 \def\ktr{{\reflectbox{\rOpe\char'170}}}        
 \def\kbr{{\reflectbox{\rOpe\char'170}}}        
 \def\Border{\vbox{\hsize0pt
        \setlength{\unitlength}{1mm}
        \newcount\xco
        \newcount\yco
        \xco=-21
        \yco=12
        \begin{picture}(0,0)(-7.5,0)
        \put(\xco,\yco){$\ktl$}
        \advance\yco by-1
        {\loop
        \put(\xco,\yco){$\kcr$}
        \advance\yco by-2
        \ifnum\yco>-240
        \repeat
        \put(\xco,\yco){$\kbl$}}
        \xco=170
        \yco=12
        \put(\xco,\yco){$\ktr$}
        \advance\yco by-1
        {\loop
        \put(\xco,\yco){$\kcr$}
        \advance\yco by-2
        \ifnum\yco>-240
        \repeat
        \put(\xco,\yco){$\kbr$}}
        \put(-19.5,13){\scalebox{.6065}{%
         University of Maryland Center for String and Particle  Theory \&\ Physics Department%
        |University of Maryland Center for String and Particle  Theory \&\ Physics Department}}
        \put(-19.5,-241.5){\scalebox{.5835}{%
         ****University of Maryland * Center for String and
         Particle  Theory* Physics Department****University of Maryland *Center
        for String and Particle  Theory* Physics Department}}
        \end{picture}
        \par\vskip-8mm}}
\definecolor{UMred}{rgb}{.9,.05,.2}
\definecolor{HUblue}{rgb}{.0,.3,.7}
 \def\UMbanner{\vbox{\hsize0pt
        \setlength{\unitlength}{.4mm}
        \thicklines  
        \begin{picture}(0,0)(-30,-10)
        \put(165,2){\line(1,0){4}}
        \put(170,2){\line(1,0){4}}
        \put(180,2){\line(1,0){4}}
        \put(175,-14){\line(1,0){4}}
        \put(180,-14){\line(1,0){4}}
        \put(185,-14){\line(1,0){4}}
        \put(169,-14){\line(0,1){16}}
        \put(170,-14){\line(0,1){16}}
        \put(179,-14){\line(0,1){16}}
        \put(180,-14){\line(0,1){16}}
        \put(184,-14){\line(0,1){16}}
        \put(185,-14){\line(0,1){16}}
        \put(169,2){\oval(8,32)[bl]}
        \put(170,2){\oval(8,32)[br]}
        \put(179,-14){\oval(8,32)[tl]}
        \put(185,-14){\oval(8,32)[tr]}
        \end{picture}
        \par\vskip-6.5mm
        \thicklines}}

\definecolor{Red}    {rgb}{0.90,0.00,0.12} 
\definecolor{Blue}   {rgb}{0.00,0.00,1.00} 
\definecolor{Green}  {rgb}{0.10,0.70,0.10} 
\definecolor{Turque} {rgb}{0.00,0.65,0.85} 
\definecolor{Orange} {rgb}{1.00,0.50,0.15} 
\definecolor{Magenta}{rgb}{1.00,0.00,1.00} 
\definecolor{Gold}   {rgb}{1.00,0.75,0.25} 
\definecolor{Seaweed}{rgb}{0.01,0.24,0.09} 
\definecolor{Purple} {rgb}{0.50,0.25,0.55} 
\definecolor{Brown}  {rgb}{0.43,0.26,0.32} 
\definecolor{grey1}  {rgb}{0.20,0.20,0.20} 
\definecolor{grey2}  {rgb}{0.40,0.40,0.40} 
\definecolor{grey3}  {rgb}{0.60,0.60,0.60} 
\definecolor{grey4}  {rgb}{0.80,0.80,0.80} 
\definecolor{grey5}  {rgb}{0.90,0.90,0.90} 
\def\C#1#2{{\ifcase#1\or
             \color{Red}\or \color{Green}\or \color{Blue}\or\
              \color{Turque}\or \color{Orange}\or \color{Magenta}\or 
               \color{Gold}\or \color{Seaweed}\or \color{Purple}\or
                \color{Brown}\or\color{grey1}\or\color{grey2}\or
                 \color{grey3}\else\color{grey4}\fi#2}}

\definecolor{Slate} {rgb}{0.00,0.45,0.55}

\newdimen\parshift\parshift=\parindent
\catcode`@=11
 \long\def\@footnotetext#1{\insert\footins{\reset@font\footnotesize
           \interlinepenalty\interfootnotelinepenalty\splittopskip%
            \footnotesep\splitmaxdepth\dp\strutbox\floatingpenalty\@MM%
             \hsize\columnwidth\addtolength{\hsize}{-2\parindent}
              \@parboxrestore\protected@edef\@currentlabel%
              {\csname p@footnote\endcsname\@thefnmark}%
                \color@begingroup%
                 \@makefntext{\rule\z@\footnotesep\ignorespaces#1%
                  \@finalstrut\strutbox}%
                \color@endgroup}}
 \long\def\@makefntext#1{\hglue\parshift%
           \vbox{\noindent\baselineskip=11pt plus.5pt minus.5pt\hb@xt@0em{\hss\@makefnmark\kern1pt}#1}}
\catcode`@=12

\newskip\humongous \humongous=0pt plus 1000pt minus 1000pt
\def\caja{\mathsurround=0pt}
\def\eqalign#1{\,\vcenter{\openup2\jot \caja
        \ialign{\strut \hfil$\displaystyle{##}$&$
        \displaystyle{{}##}$\hfil\crcr#1\crcr}}\,}
\newif\ifdtup
\def\panorama{\global\dtuptrue \openup2\jot \caja
        \everycr{\noalign{\ifdtup \global\dtupfalse
        \vskip-\lineskiplimit \vskip\normallineskiplimit
        \else \penalty\interdisplaylinepenalty \fi}}}

\makeatletter
\def\section{\@startsection{section}{1}{\z@}
        {3ex plus-1ex minus-.2ex}{1pt plus1pt}{\large\sf\bfseries\boldmath}}
\def\subsection{\@startsection{subsection}{2}{\z@}
         {1.5ex plus-1ex minus-.2ex}{0.01pt plus1pt}{\sf\slshape}}
\def\subsubsection{\@startsection{subsubsection}{3}{\z@}
          {1.5ex plus-1ex minus-.2ex}{0.01pt plus0.2pt}{\sf\boldmath}}
\def\paragraph{\@startsection{paragraph}{4}{\z@}
           {.75ex \@plus.5ex \@minus.2ex}{-2mm}{\sf\bfseries\boldmath}}
\makeatother

 \allowdisplaybreaks
 \seceq

\begin{document}

\thispagestyle{empty}
\vbox{\Border\UMbanner}
 \noindent{\small
 \today\hfill{PP-016-001 
 }}
  \vspace*{8mm}
 \begin{center}
{\large \bf
A Proposal On Culling \& Filtering A Coxeter Group   \\[2mm]
 For  4D, $\bm {\cal N}$ = 1
Spacetime SUSY Representations}
 \\[12mm]
{\large D.\ E.\ A.\  Gates\footnote{dgates@g.harvard.edu}${}^{\dag}$,~ and\,
                    S.\ James Gates, Jr.$^*$   $^{\#}$\footnote{gatess@wam.umd.edu}
                    }\\*[12mm]
\emph{
\centering
$^*$Center for String and Particle Theory, Dept.\ of Physics, \\[-2pt]
University of Maryland, College Park, MD 20472,  USA
\\[12pt]      
$^{\#}$Department of Physics and Astronomy \\[-2pt]
Dartmouth College, NH 03755, USA 
\\[6pt]
and
\\[6pt]
$^\dag$ Jefferson Physical Laboratory, Harvard University 
\\[1pt]
Cambridge, MA 02138, USA
}
 \\*[40mm]
{ ABSTRACT}\\[4mm]
\parbox{142mm}{\parindent=2pc\indent\baselineskip=14pt plus1pt
We review the mathematical tools required to cull and filter representations
of the Coxeter Group $BC_4$ into providing bases for the construction of 
minimal off-shell representations of the 4D, $ {\cal N}$ = 1 spacetime 
supersymmetry algebra.  Of necessity this includes a description of the 
mathematical mechanism by which four dimensional Lorentz symmetry 
appears as an emergent symmetry in 
the context of one dimensional adinkras with four colors described by 
the Coxeter Group $BC_4$.
 }
 \end{center}
\vfill
\vfill
\clearpage

\section{Introduction}

Recently, one of us (D.E.A.G.) gave a presentation at the 2015 Miami Topical 
Physics Conference and during a question-and-answer session afterward there
arose a query from Prof.\ J.\ Lukierski about the non-obvious relationship 
between the Euclidean SO(4) symmetry manifest in the use of adinkras \cite{adnk1} 
and their adjacency matrices \cite{GRana,GrphThry} versus the Lorentzian structure required to 
describe the usual theories of interest realizing the 4D, ${ {\cal N}}$ = 1 
spacetime supersymmetry algebra.

This exchange motivates us to review tools developed in previous works
to show evidence that although adinkras with four colors, four open nodes, and
four closed nodes manifestly realize a Euclidean SO(4) symmetry there is a
 ``hidden path'' that also relates each of them to a set of SO(1,3) Dirac $\bm \g$-matrices.

First, we review results that are standard to the usual Dirac matrices 
appropriate for a four dimensional Minkowski space.  We also show how the 
generators of SO(4) possess an apparently often overlooked relationship to 
the Dirac matrices appropriate for a four dimensional Minkowski space.  There
follows a discussion of the L-matrices and R-matrices \cite{GRana} associated 
with every adinkra graph \cite{adnk1} obtainable by a modification 
of the standard notion of an adjacency graph \cite{GrphThry}.  The commutator 
algebra of the L-matrices and R-matrices define the `holoraumy' 
\cite{KIAS,HoLoRmY,HoLoRmY4D} associated with the graph.

In this presentation, we restrict ourselves to the case of adinkras with four colors, 
four open nodes, and four closed nodes because we constructed all possible 
representations of this kind in a previous analysis \cite{{permutadnk}}. Using a 
computer software program, it was found one can start with the Coxeter 
Group $BC_4$ and associate with every element of this group to some adinkra.  
In particular, quartets of elements of  $BC_4$ form representations of the adjacency 
matrices associated with this class of adinkras.  This previous work showed 
there are 1,536 such quartets.  By taking the elements of the $BC_4$ Coxeter 
Group as our starting point, we have a rigorous mathematically well-defined 
beginning for our analysis. The fifth section presents the criteria by which a subset 
of the elements of Coxeter Group $BC_4$ can be consistently interpreted as a 
projection of the 4D, $\cal N$ = 1 fundamental irreducible supermultiplet representations.  
Thus we also identify obstructions that prevent such identifications for all the elements 
of $BC_4$. 

\section{Connecting Dirac SO(1,3) $\bm \g$-Matrices To SO(4) Rotation Matrices}
\label{s3}

A set of Dirac $\g$-matrices is provided by ${\bm {\g}}^{\mu}$ $=$ $(\, {\bm {\g}}^{
0} , \, {\bm {\g}}^{1} , \, {\bm {\g}}^{2} , \, {\bm {\g}}^{3}  \,)$ and must satisfy the usual 
condition
\be
{\bm {\g}}^{\mu} \, {\bm {\g}}^{\nu} ~+~ {\bm {\g}}^{\nu} \, {\bm {\g}}^{\mu}  
~=~ 2\, \eta^{\m \, \n}   \, {\bf I}_4   ~~~,
\label{GammaM2}
\ee
where the 4 $\times$ 4 identity matrix is denoted by ${\bf I}_4$ and the Minkowski 
metric $\eta^{\m \, \n}$ in (\ref{GammaM2}) has non-vanishing diagonal entries
$(\, -1, \, + 1,  \,+ 1,  \,+ 1 \,) $.

Given a set of Dirac gamma matrices ${\bm {\g}}^{\mu}$, we define ${\bm {\g}}^{5}$
via the usual definition
\be
 {\bm {\g}}^{5}  ~=~ i \, {\bm {\g}}^{0}  {\bm {\g}}^{1}  {\bm {\g}}^{2}  {\bm {\g}}^{3}
 ~~~,
 \label{GammaM4}
\ee
and define a representation of the generators of spatial rotations provided  by the set  
containing the three elements $\bm \{$
${\bm \sigma}^{1 \, 2}$, ${\bm \sigma}^{2 \, 3}$, ${\bm \sigma}^{3 \, 1}$ $\bm \}$
where
\be
{\bm \sigma}^{1 \, 2} ~=~-\, i \,  {\bm {\g}}^{1}  {\bm {\g}}^{2}  ~~,~~ 
{\bm \sigma}^{2 \, 3} ~=~- \, i \,  {\bm {\g}}^{2}  {\bm {\g}}^{3}  ~~,~~ 
{\bm \sigma}^{3 \, 1} ~=~ - \, i \,   {\bm {\g}}^{3}  {\bm {\g}}^{1}  ~~.
 \label{GammaM6}
\ee
The commutator algebra of these takes the usual form
\be{
[ \,   {\bm \sigma}^{1 \, 2}  ~,~ {\bm \sigma}^{2 \, 3}   \, ] ~=~ i \, 2 \, {\bm \sigma}^{3 \, 1}  ~~,~~
[ \,   {\bm \sigma}^{2 \, 3}  ~,~ {\bm \sigma}^{3 \, 1}   \, ] ~=~ i \, 2\, {\bm \sigma}^{1 \, 2}  ~~,~~
[ \,   {\bm \sigma}^{3 \, 1}  ~,~ {\bm \sigma}^{1 \, 2}   \, ] ~=~ i \, 2\, {\bm \sigma}^{2 \, 3}  ~~,
 \label{GammaM7a}
}\ee
with all other commutators vanishing.

We now introduce another set of matrices containing three elements $\bm \{$
$i\, {\bm {\g}}^{0}$, $ {\bm {\g}}^{5}$, $ {\bm {\g}}^{0}  {\bm {\g}}^{5}$ $\bm \}$.  The
commutator algebra of these elements is
\be{
[ \, i\,  {\bm {\g}}^{0}  ~,~ {\bm {\g}}^{5}   \, ] ~=~ i \, 2\,   {\bm {\g}}^{0}  {\bm {\g}}^{5} ~~,~~
[ \,   {\bm {\g}}^{5}  ~,~ {\bm {\g}}^{0} {\bm {\g}}^{5}   \, ] ~=~ i \, 2 \, (i \,  {\bm {\g}}^{0}  )~~,~~
[ \, {\bm {\g}}^{0}  {\bm {\g}}^{5}  ~,~ i \, {\bm {\g}}^{0}   \, ] ~=~ i \, 2\, (  {\bm {\g}}^{5}  )
  ~~,
   \label{GammaM8a}
}\ee
with all other commutators vanishing.  The form of this commutator algebra
(\ref{GammaM8a}) is identical to the one in (\ref{GammaM7a}) and both are 
recognizable as SU(2) algebras.  Furthermore, it is easy to show
\be{
[ \, i\,  {\bm {\g}}^{0}  ~,~ {\bm \sigma}^{i \, j}   \, ] ~=~0 ~~,~~
[ \,   {\bm {\g}}^{5}  ~,~   {\bm \sigma}^{i \, j}    \, ] ~=~0 ~~,~~
[ \, {\bm {\g}}^{0}  {\bm {\g}}^{5}  ~,~  {\bm \sigma}^{i \, j}  \, ] ~=~ 0
  ~~.
   \label{GammaM9a}
}\ee
This all implies given a set of  Dirac $\bm \g$-matrices together with a ${\bm \g}^{5} $-matrix 
it is possible to construct the six matrices above (\ref{GammaM6}) and  (\ref{GammaM8a}), 
each forming a representation of an SU(2) algebra,  and each SU(2) algebra 
commutes with the other.

A standard result for $\bm {\g}$-matrices implies
\be   {
 {\bm {\g}}^1 ~=~  {\bm {\g}}^0  {\bm {\g}}^5 \, {\bm \sigma}^{2 \, 3}
 ~~,~~
  {\bm {\g}}^2 ~=~  {\bm {\g}}^0  {\bm {\g}}^5 \, {\bm \sigma}^{3 \, 1}
 ~~,~~
{\bm {\g}}^3 ~=~  {\bm {\g}}^0  {\bm {\g}}^5 \, {\bm \sigma}^{1 \, 2}
~~~,   } \label{GMMA}
\ee
which demonstrates that given the data of the two distinct SU(2) matrices, the
three spatial $\bm \g$-matrices can be reconstructed.  Actually, the information in both 
commuting SU(2)'s is over-complete as it is only the ``third'' component of
the ``non-orbital'' SU(2) along the three components of the ``orbital'' SU(2)
that are required. 

At this point, a different set of 4 $\times$ 4 matrices can be introduced via 
the definitions
\be
\eqalign{
{~~} {\bm {\a}}^1  \,&=\, {\bm \s}^2 \otimes {\bm \s}^1   ~\,~~,~~
{\bm {\b}}^1   \,
=\, {\bm \s}^1 \otimes {\bm \s}^2 ~ \,\,, \cr
{\bm {\a}}^2  \,&=\, {\bf I}  \otimes {\bm \s}^2   ~~~~~~,~~
{\bm {\b}}^2  \,=\, {\bm \s}^2 \otimes {\bf I} ~~\, \,\,,   \cr  
{\bm {\a}}^3  \,&=\, {\bm \s}^2 \otimes {\bm \s}^3 ~\,~~,~~
{\bm {\b}}^3 \,=\,  {\bm \s}^3 \otimes {\bm \s}^2 \,\, \,\,,
}
\label{alpbet}
\ee
where these matrices satisfy the identities
\be  \eqalign{ {~~~~~~~}
{\bm \a}^{\Hat I} \, {\bm \a}^{\Hat K} ~&=~ \delta{}^{{\Hat I} \, {\Hat K}} \, {\bm {\rm I}}{}_4
~+~ i \, \epsilon{}^{{\Hat I}  \, {\Hat K} \, {\Hat L}} \, {\bm \a}^{\Hat K} ~~~, \cr
{\bm \b}^{\Hat I} \, {\bm \b}^{\Hat K} ~&=~ \delta{}^{{\Hat I} \, {\Hat K}} \,  {\bm {\rm I}}{}_4 
~+~ i \, \epsilon{}^{{\Hat I}  \, {\Hat K} \, {\Hat L}} \, {\bm \b}^{\Hat K} 
~~~,~~   \cr
[\,  {\bm \a}^{\Hat I}   ~,~ {\bm \b}^{\Hat J}  \,  ] ~&=~ 0 ~~.
 }   \label{Trcs}
\ee
The commutator algebra derivable from (\ref{Trcs}) allows us to identify the six matrices 
(\ref{alpbet}) as the hermitian 4$\times$4 matrix generators of SO(4).  We also have
\be  \eqalign{ {~~~~~~~}
& {~} {\rm {Tr}} \big( \,  {\bm \a}^{\Hat I}  \, {\bm \a}^{\Hat J}  \, \big) ~=~ {\rm {Tr}} \big( \, 
{\bm \b}^{\Hat I}  \,  {\bm \b}^{\Hat J} \, \big) ~=~ 4\,  \delta{}^{{\Hat I} \, {\Hat J}} ~~,~~ 
\ {\rm {Tr}} \big( \,  {\bm \a}^{\Hat I}  \, {\bm \b}^{\Hat J}  \, \big) 
~=~ 0 ~~,   \cr
& {~~~~~~~~~~~~~~~~~~~~~} {\rm {Tr}} \big( \,  {\bm \a}^{\Hat I}   \, \big) ~=~ {\rm {Tr}} \big( \, 
{\bm \b}^{\Hat I}  \,  \, \big) ~=~ 0 ~~~.
}   \label{Trcs2}
\ee

However, the commutator algebra defined by (\ref{GammaM7a}), (\ref{GammaM8a}), 
and (\ref{GammaM9a}) is isomorphic to that which is derivable from  (\ref{Trcs}).  Hence
both are representations of SO(4).  Therefore, the Dirac gamma matrices can be 
expressed using the `$\bm \a$-set' and `$\bm \b$-set'.  One such set of definitions are
 \be
\eqalign{
{~~} {\bm {\g}}^0  \,&=\, i\, {\bm \b}^3   ~~\,~,~~~~
{\bm {\g}}^1   \,
=\, {\bm \a}^1  {\bm \b}^2    ~~,~~~~
{\bm {\g}}^2  \,=\, {\bm \a}^2  {\bm \b}^2  ~~\,\,~~,~~~~
{\bm {\g}}^3 \,=\,  {\bm \a}^3  {\bm \b}^2 
 \,\, \,\,,
}
\label{alpbet2}
\ee
which imply
\be
{\bm \sigma}^{1 \, 2} ~=~    {\bm {\a}}^{3}  ~~,~~ 
{\bm \sigma}^{2 \, 3} ~=~  {\bm {\a}}^{1}   ~~,~~ 
{\bm \sigma}^{3 \, 1} ~=~  {\bm {\a}}^{2}  ~~.
 \label{GammaM7}
\ee
These equations make manifest that the ${\bm \sigma}{}^{i \, j}$-matrices are the
generators of SU(2).   A definition of the ${\bm {\g}}^5 $ matrix chosen as
 \be
 {\bm {\g}}^5 ~=~ - \, {\bm {\b}}^1  ~~~,
 \label{GammaM8}
\ee
together with the definition of  ${\bm {\g}}^0$ shown above implies
 \be
\eqalign{
{\bm {\g}}^0  {\bm {\g}}^5 \,=\, \, {\bm \b}^2  ~~\,\,~~.
}
\label{GammaM9b}
\ee
It is not commonly noted that given a set of four dimensional gamma matrices, it is
possible to use them to define {\em {two}} {\em {mutually}} {\em {commuting}} 
SU(2)'s and this is valid independent of the representation chosen for the 
gamma matrices.   

Let us also note the identifications between the $\bm \g$-matrices on the one side 
of the equations (\ref{alpbet2}) and (\ref{GammaM7}) and the $\bm \a$-matrices 
and the $\bm \b$-matrices on the other are not unique.  One can cyclically permute, 
independently, the $\bm \a$-matrices and the $\bm \b$-matrices in these equation 
and this still leads to a properly defined set of $\bm \g$-matrices.  Similarly, one
can perform the exchanges of the form ${\bm \a}^{\Hat I} \leftrightarrow {\bm \b}^{
\Hat I}$  simultaneously in the equations (\ref{alpbet2}) and (\ref{GammaM7}) 
and this also leads to a properly defined set of $\bm \g$-matrices.  It is an open
question to ask, ``What is the group of transformations acting on the $\bm \a$-matrices 
and the $\bm \b$-matrices in (\ref{alpbet2}) such that the so defined $\bm \g$-matrices
satisfy (\ref{GammaM2})?''  The bottom line is there is {\it {not}} a unique way for 
a set of Dirac $\bm \g$-matrices for a four dimensional Minkowski space with 
Lorentz symmetry to emerge from the two commuting SU(2) groups constructed 
from SO(4).

However, the fact that SO(4) can `secretly' carry information about SO(1,3) spinors 
and $\bm {\g}$-matrices is one of the important mechanisms for use of adinkras 
with four colors to describe 4D, $\cal N$ = 1 SUSY theories that utilize Minkowski 
space spinors.

\section{From L-matrices, R-matrices to Dirac $\bm \g$-matrices}
\label{s3B}
In this section, we will show how the L-matrices and R-matrices \cite{GRana}
that occur in the description of any adinkra graph \cite{adnk1}, \cite{GrphThry}
with four colors (${\rm I} = 1,\dots, 4$) four open nodes ($i = 1,\dots, 4$), and 
four closed nodes  (${\hat k} = 1,\dots, 4$) are related to a set of SO(4) rotation 
matrices.  From the last section, we showed there exist a possibility of linking 
the $\bm \g$-matrices of SO(1,3) to the SO(4) rotation matrices.  Combining 
these two results, we thus find a pathway that connects all adinkras with four 
colors, four open nodes, and four closed nodes to the representations of 
SO(1,3) $\bm \g$-matrices.

Every adinkra representation $(\cal R)$ leads to a set of four adjacency matrices 
denoted by ${\rm L}^{(\cal R)}_\rI\,$ and ${\rm R}^{(\cal R)}_\rI$ which satisfy
the conditions
\be { \eqalign{
 (\,{\rm L}^{(\cal R)}_\rI\,)_i{}^\hj\>(\,{\rm R}^{(\cal R)}_\rJ\,)_\hj{}^k ~+~ (\,{\rm L
 }^{(\cal R)}_\rJ\,)_i{}^\hj\>(\,{\rm R}^{(\cal R)}_\rI\,)_\hj{}^k &= 2\,\d_{\rI\rJ}\,\d_i{}^k~~,\cr
 (\,{\rm R}^{(\cal R)}_\rJ\,)_\hi{}^j\>(\, {\rm L}^{(\cal R)}_\rI\,)_j{}^\hk ~+~ (\,{\rm 
 R}^{(\cal R)}_\rI\,)_\hi{}^j\>(\,{\rm L}^{(\cal R)}_\rJ\,)_j{}^\hk  &= 2\,\d_{\rI\rJ}\,\d_\hi{
 }^\hk~~,  \cr
~~~(\,{\rm R}^{(\cal R)}_\rI\,)_\hj{}^k\,\d_{ik} = (\,{\rm L}^{(\cal R)}_\rI\,)_i{
}^\hk\,\d_{\hj\hk}&~~.
}}\label{GarDNAlg2}
\ee
This we call the ``Garden Algebra.''
Given a set of L-matrices and R-matrices for a specified adinkra representation
$(\cal R)$, we can define two additional matrix sets denoted by $\bm {V_{\rI\rJ
}^{(\cal R)}}$ and $\bm{{\Tilde V}_{\rI\rJ}^{(\cal R)}}$ \cite{KIAS,HoLoRmY,HoLoRmY4D}
via the equations
\be { \eqalign{
 (\,{\rm L}^{(\cal R)}_\rI\,)_i{}^\hj\>(\,{\rm R}^{(\cal R)}_\rJ\,)_\hj{}^k ~-~ (\,{\rm L
 }^{(\cal R)}_\rJ\,)_i{}^\hj\>(\,{\rm R}^{(\cal R)}_\rI\,)_\hj{}^k &= i\, 2\,  (V_{\rI\rJ}^{(\cal 
 R)})_i{}^k~~,\cr
 (\,{\rm R}^{(\cal R)}_\rJ\,)_\hi{}^j\>(\, {\rm L}^{(\cal R)}_\rI\,)_j{}^\hk ~-~ (\,{\rm 
 R}^{(\cal R)}_\rI\,)_\hi{}^j\>(\,{\rm L}^{(\cal R)}_\rJ\,)_j{}^\hk  &= i \, 2\,  
 ({\Tilde V}_{\rI\rJ}^{(\cal R)})_\hi{}^\hk ~~.
}}\label{GarDVs}
\ee
We have given the name of ``bosonic holoraumy matrices'' to the quantities $\bm {V_{\rI
\rJ}^{(\cal R)}}$ and  ``fermionic holoraumy matrices'' to the quantities $\bm{{\Tilde V}_{\rI
\rJ}^{(\cal R)}}$ defined here.  Due to the definitions in (\ref{GarDNAlg2}), it follows that both 
sets of holoraumy matrices satisfy the commutator algebra that describes SO(4).  Since the 
$\bm{{\Tilde V}_{\rI\rJ}^{(\cal R)}}$ matrices act in the spinor space of the adinkras, we 
concentrate upon it.  This means we can write an equation of the form
\be   \eqalign{
 {\bm {\Tilde V}_{\rI\rJ}^{(\cal R)}} ~=~ &\Big[ \,\ell^{({\cal R})1}_{\rI\rJ}\, {\bm 
 {\a^1}} \, + \,  \ell^{({\cal R})2}_{\rI\rJ}\, {\bm {\a^2}}  \,+\,  \ell^{({\cal R})3
 }_{\rI\rJ}\, {\bm {\a^3}}   \, \Big]      \cr
 &~+~  \Big[ \,    {{\Tilde \ell}^{(\cal R)}}_{\rI\rJ}{
 }^{1}\, {\bm {\b^1}}  \,+\, \, {{\Tilde \ell}^{(\cal R)}}_{\rI\rJ}{}^{2}\, {\bm {\b^2}}   
 \,+\, {{\Tilde \ell}^{(\cal R)}}_{\rI\rJ}{}^{3}\, {\bm {\b^3}}   \, \Big]   ~~~,
}  \label{Veq}
\ee
for some set of coefficients $\ell^{({\cal R})1}_{\rI\rJ}$, $\ell^{({\cal R})2}_{\rI\rJ}$,
$\ell^{({\cal R})3}_{\rI\rJ}$, ${\Tilde \ell}^{({\cal R})1}_{\rI\rJ}$, ${\Tilde \ell}^{({\cal 
R})2}_{\rI\rJ}$, and ${\Tilde \ell}^{({\cal R})3}_{\rI\rJ}$.
Using the results of the last chapter, this becomes
\be
  \eqalign{
 {\bm {\Tilde V}_{\rI\rJ}^{(\cal R)}} ~=~   &\Big[ \,\ell^{({\cal R})1}_{\rI\rJ}\, {\bm 
 {\S^{23}}} \, + \,  \ell^{({\cal R})2}_{\rI\rJ}\, {\bm {\S^{31}}}  \,+\,  \ell^{({\cal R})3
 }_{\rI\rJ}\, {\bm {\S^{12}}}   \, \Big]   \cr
 &~+~  \Big[ \,  -\,   {{\Tilde \ell}^{(\cal R)}}_{\rI\rJ}{
 }^{1}\, {\bm {\g^5}}  \,+\, \, {{\Tilde \ell}^{(\cal R)}}_{\rI\rJ}{}^{2}\, {\bm {\g^0 \g^5}}   
 \,+\, i \,  {{\Tilde \ell}^{(\cal R)}}_{\rI\rJ}{}^{3}\, {\bm {\g^0}}   \, \Big]    ~~~.
}  \label{Veq2}
\ee
We have referred to (\ref{Veq2}) in the past \cite{KIAS} as the ``Adinkra/$\bm \g$-matrix Holography 
Equation.''

The importance of (\ref{Veq2}) when combined with  (\ref{GMMA}) is that it implies
that for any four color, four open-node, four-closed node adinkra along with the
introduction of the complete specification of two distinct commuting SU(2) algebras,
$\{ \S^{i \, j} \}$ and $\{ i\, {\bm {\g}}^{0}, \, {\bm {\g}}^{5}\, ,  {\bm {\g}}^{0}  {\bm {\g}}^{5} \}$,
derivable from adinkras, it is possible to find a set of three spatial $\gamma$-matrices
and connect to the Lorentz symmetries.  The link between any specific adinkra, of the 
type under consideration, to the representations of the Minkowski space SU(2) algebras,
$\{ \S^{i \, j} \}$ and $\{ i\, {\bm {\g}}^{0}, \, {\bm {\g}}^{5}\, ,  {\bm {\g}}^{0}  {\bm {\g}}^{5} \}$,
is specified by the constants $\ell^{({\cal R})1}_{\rI\rJ}$, $\ell^{({\cal R})2}_{\rI\rJ}$,
$\ell^{({\cal R})3}_{\rI\rJ}$, ${\Tilde \ell}^{({\cal R})1}_{\rI\rJ}$, ${\Tilde \ell}^{({\cal 
R})2}_{\rI\rJ}$, and ${\Tilde \ell}^{({\cal R})3}_{\rI\rJ}$.

\section{The Coxeter Group $\bm {BC_4}$ Embedding Starting Point}
\label{s4}

For our purposes, we can define the elements of $BC_4$ \cite{CXgrp} in the 
following manner.  Consider the set of all real 4 $\times$ 4 matrices of the form
\cite{permutadnk}
\begin{equation}
 {\bm \rL} ~=~ 
     {\bm {\cal S}} \, {\bm {\cal P}}
\label{aas0}
\end{equation}
We call the matrix ${\bm {\cal S}}$  the ``Boolean Factor''  \cite{permutadnk}
as it is a real
diagonal $4\times4$ matrix that squares to the identity.  The matrix ${\bm {
\cal P}}$ is a matrix representation of a permutation of $4$ objects. There 
are $ 2^d \, d!$ =  $4! \times 2^4=384$ ways to choose the Boolean Factor 
{\em {and}} the Permutation matrix. This is the dimension of the Coxeter 
group $BC_4$.

By embedding the L-matrices as the elements in the entirety of $BC_4$ we
know that for each one we can write the equation
\begin{equation}
 (\rL_{\sss\rI} {}^{({\cal R})})_i{}^\hk ~=~ 
     [{\cal S}^{\sss(\rI)} {}^{({\cal R})}]_i{}^\hl\, [{\cal P}_{\!\sss(\rI)} {}^{({\cal R})}]_\hl{}^\hk,
      \qquad \text{for each fixed }\rI=1,2,\dots,N.
\label{aas1}
\end{equation}
This notation anticipates that there are distinct adinkra representations
denoted by the label $({\cal R})$ and each adinkra leads to four matrices
labeled by the index $\rm I$.
In other words,
the L-matrix for a single fixed value of I can be chosen to be any element in the
Coxeter group $BC_4$.

So if we were simply picking quartets of distinct elements of the Coxeter group $BC_4$ in
an arbitrary manner there would be $n_4$ where 
\be
n_4 ~=~ \frac{384 \cdot 383 \cdot 382 \cdot 381} {4 !} ~=~  891,881,376
\label{4-Ways}
 \ee
ways to pick the elements.  However, we wish to pick the distinct four elements 
of the $BC_4$ Coexeter Group so that they satisfy the ``Garden Algebra.''  This 
requirement is so severe there are only 1,536 ways in which the four elements 
of the $BC_4$ Coexeter Group can be chosen to form a supersymmetry quartet.  
This was discovered by utilizing a code \cite{permutadnk} to exhaustively construct 
all possible quartets starting from the  $BC_4$ Coexeter Group elements.  The label
$({\cal R})$ written in (\ref{aas1}) takes its values over these representations.

This startlingly smaller number is mostly determined by the permutation elements
from which any L-matrix is constructed.  It turns out only particular choices of
the permutation elements can appear within any given quartet.  This is shown in
the following collections of sets
\be  \eqalign{
{\bm \{} {\cal P}_1  {\bm \} }    ~=~&{\bm \{}  (123), ~ (134 ), ~ (142), ~ (243) {\bm \} }  ~=~  
 (123)\, {\bm \{} {\cal V}  {\bm \} }  ~~~, \cr
{\bm \{} {\cal P}_2  {\bm \} }    ~=~&{\bm \{}  (124), ~ (132 ), ~ (143), ~(234) {\bm \} }  
 ~=~   (124)\, {\bm \{} {\cal V}  {\bm \} }  ~~~, \cr
{\bm \{} {\cal P}_3  {\bm \} }    ~=~&{\bm \{}  (14), ~ (23 ), ~(1243), ~ (1342) {\bm \} }    
 ~=~   (14)\, {\bm \{} {\cal V}  {\bm \} } ~~~, \cr
{\bm \{} {\cal P}_4  {\bm \} }    ~=~&{\bm \{} (13) , ~ (24), ~ (1234), ~  (1432){\bm \} }  
 ~=~   (13)\, {\bm \{} {\cal V}  {\bm \} }   ~~~, \cr
{\bm \{} {\cal P}_5  {\bm \} }    ~=~&{\bm \{}  (12), ~ (34), ~ (1243),~ (1324) {\bm \} } 
 ~=~   (12)\, {\bm \{} {\cal V}  {\bm \} }    ~~~, \cr
{\bm \{} {\cal P}_6  {\bm \} }    ~=~&{\bm \{} (), ~ (12)(34), ~ (13)(24), ~ (14)(23) {\bm \} }   
 ~=~   {\bm \{} {\cal V}  {\bm \} }  ~~~, \cr
} \label{PermSets}
\ee
where we have used cycle notation to indicate the distinct permutations and relate all
the permutations to the Vierergruppe\footnote{We thank our colleague K.\ Iga for this observation.}
denoted by $ {\bm \{} {\cal V}  {\bm \} } $ \cite{4Grp} thus making manifest its
critical role.

\begin{figure}[ht]
\begin{center}
\begin{picture}(70,60)
\put(-16,-06){\includegraphics[height = 67\unitlength]{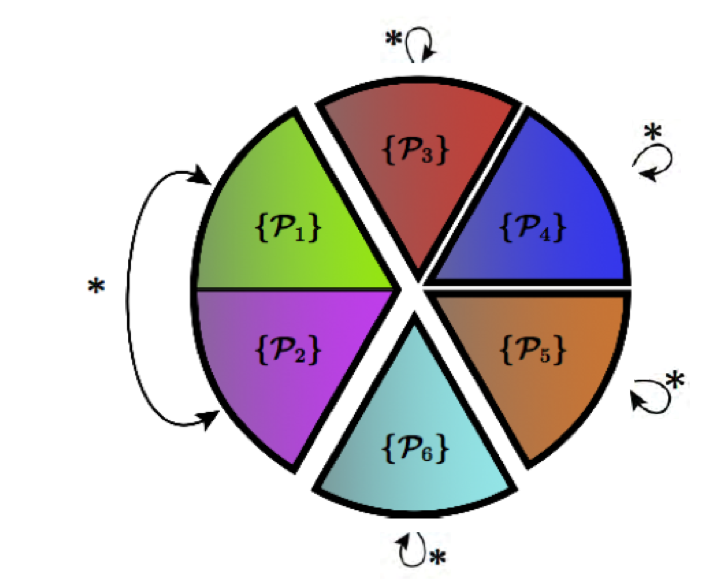}}
\end{picture}
\end{center}
\caption{Orbit space of ${\cal P}$ permutation matrices under the $*$ map}
\label{f:pie}
\end{figure}

The action of transposition (denoted by the symbol ${}^{\bm *}$) on these sets 
is straightforward to calculate and we find
\begin{equation}
 \begin{array}{r@{\>=\>}l}
  {}^{\bm*}\{   {\cal P}_1  \}&\{  {\cal P}_2  \}~,\\
  {}^{\bm*}\{   {\cal P}_2   \}&\{   {\cal P}_1  \}~,\\
 \end{array}
 \qquad
  \begin{array}{r@{\>=\>}l}
  {}^{\bm*}\{   {\cal P}_3  \}&\{  {\cal P}_3  \}~,\\
  {}^{\bm*}\{   {\cal P}_4   \}&\{   {\cal P}_4  \}~,\\
 \end{array}
 \qquad
   \begin{array}{r@{\>=\>}l}
  {}^{\bm*}\{   {\cal P}_5  \}&\{  {\cal P}_5  \}~,\\
  {}^{\bm*}\{   {\cal P}_6   \}&\{   {\cal P}_6  \}~,\\
 \end{array}
 \qquad
  \label{HodgeSTAR}
\end{equation}
and in writing this, we define two sets to be the same if they contain the
same elements irrespective of the order in which they appear.  In Figure
(\ref{f:pie}) these subsets of permutations together with the action of the $\bm *$
map are shown.
 
\section{Culling \& Filtering}
\label{C&F}

In a previous work \cite{BwT1}, there was presented an obstruction that 
indicated when an adinkra with two colors was compatible with being the 
projection of a two dimensional supermultiplet.  It was shown there exist 
what we may call the ``no two-color ambidextrous bow tie'' theorem which 
asserted if an adinkra graph contained a certain structure, then it was not 
possible to consistently ``lift'' the adinkra graph in such a way that it could 
be associated with a supermultiplet defined on a Minkowski space with a 
Lorentzian metric with diagonal entries of the (-1,1) variety.  

Up until now we have made no similar comments about when an arbitrary 
four-color adinkra can be regarded as being associated with the projection 
of a supermultiplet defined on a Minkowski space with a Lorentzian metric 
with diagonal entries of the (-1, 1, 1, 1) variety.   Due to our analysis of 
$BC_4$, we now have enough hints so as be comfortable laying out a 
set of analogous requirements for all adinkras based on $BC_4$ elements.  
We are now ready to describe how a subset of the elements of $BC_4$ 
can be used to construct the off-shell minimal representations of 4D, 
$\cal N$ = 1 SUSY. 

This chapter will depend crucially on some conjectures for which we
do not have closed-form explicit mathematical proofs.

$~~~$ {\it {Conjecture \# 1}}:  \newline \indent $~~~$
{\it {Given an arbitrary element in}} $BC_4$ {\it {it is always possible to
find three additional  }} \newline \indent $~~~$
{\it {distinct elements so that this quartet of distinct elements satisfies the ``Garden 
}} \newline \indent $~~~$
{\it {Algebra'' with four colors.}}
\newline $~~$ \newline \indent $~~~$
{\it {Conjecture \# 2}}:  \newline \indent $~~~$
{\it {Given an a quartet of elements in}} $BC_4$ {\it {that satisfies the ``Garden 
Algebra'' }} \newline \indent $~~~$
{\it {with four colors, their holoramy tensors always takes the form given 
in (\ref{Veq}) }} \newline \indent  $~~~$
{\it {and (\ref{Veq2}) with either all of the }} $\ell$-{\it{coefficients equal 
to zero or all of the }} $\Tilde \ell$-{\it {}} \newline \indent  $~~~$
{\it {coefficients equal to zero. }} 
\vskip,2in

Although we do not have a closed-form explicit mathematical proof of the 
first conjecture, the explicit construction in the work of \cite{permutadnk} gives
us confidence in its validity.  The code described therein constitutes a
proof by exhaustive examination.  For the second conjecture, we can make 
no such claim.  We have examined some number of specific cases and these 
all indicated such a result holds.  Under the assumption of the correctness 
of the two conjectures, the process of culling and filtering of the elements 
of $BC_4$ to consistently describe 4D, $ {\cal N}$ = 1 spacetime supermultiplets 
only requires the application of the tools of the $\bm *$-map and the holoraumy 
tensor ${\bm {\Tilde V}_{\rI\rJ}^{(\cal R)}}$.

One can pick an arbitrary element of $BC_4$ and examine how it
behaves with respect to the $\bm *$-map acting on the permutation 
upon which the element is constructed.  The permutation associated 
with the element will be in one of the ``even'' sets ($\{ {\cal P}_3  \}$ 
thru $\{ {\cal P}_6  \}$) or one of the ``odd'' sets  ($\{ {\cal P}_1  \}$ or $\{ 
{\cal P}_2  \}$).  If the permutation associated with the element is in 
one of the ``even'' sets, we next calculate the holoraumy associated 
with it. Let us call the element our base element.  By conjecture (\# 2) 
this must take to the form of (\ref{Veq}) with half of the coefficients vanishing.

Now there comes a subtlety.  In going from (\ref{Veq}) to (\ref{Veq2})
there is an ambiguity.  To go from the former to the latter required
the identifications made in (\ref{alpbet2}) and (\ref{GammaM7}).
However, as we discussed below the latter equations, there is always
an inherent ambiguity as identified in the discussion above (\ref{GMMA}).  
So using this ambiguity we can simply declare that whatever explicit 
matrices emerge from the holoraumy associated with this base element 
are associated with the orbital SU(2).

To the skeptical reader on this point, we should also note this also 
emphasizes that the 4D Lorentz symmetry is an emergent symmetry.  Before 
the choice of which adinkra based SU(2) symmetry corresponds to the 
orbital SU(2) of Minkowski space, both adinkra based SU(2) symmetry 
groups are equivalent.

This choice immediately culls and filters the rest of the $BC_4$
elements dependent on even permutations.  Given a second
element dependent upon an even permutation, if its holoraumy 
tensor commutes with that of the base element, this second element 
does not provide an example that can be reached by projection of any 
4D, $ {\cal N}$ = 1 spacetime supermultiplet.  On the other hand,  given 
a second element dependent upon an even permutation, if its holoraumy 
tensor does not commute with that of the base element, this second 
element does provide an example that can be reached by projection 
of a 4D, $ {\cal N}$ = 1 spacetime supermultiplet.

The process we have described above provides a set theoretic
definition of a 4D, $ {\cal N}$ = 1 spacetime vector supermultiplet
based solely on the properties of elements of $BC_4$.  We must
still consider the $BC_4$ elements that depend on odd permutations.

For the $BC_4$ elements dependent on odd permutations, we
now imagine calculating the holoraumy tensors.  Given conjecture
(\# 2), some of these will have holoraumy tensors that commute
with the vector supermultiplet holoraumy tensor as defined above.  
Others will have  holoraumy tensors that do not commute
with the vector supermultiplet holoraumy tensor as defined above. 

 If the $BC_4$ elements dependent on odd permutations possess
 holoraumy tensors that commute with the vector supermultiplet 
 holoraumy tensor as defined above, then such elements describe
 the projections of  4D, $ {\cal N}$ = 1 spacetime chiral supermultiplets.
 
  If the $BC_4$ elements dependent on odd permutations possess
 holoraumy tensors that do not commute with the vector supermultiplet 
 holoraumy tensor as defined above, then such elements describe
 the projections of  4D, $ {\cal N}$ = 1 spacetime tensor supermultiplets.
 
 Notice that the definitions given above depend only on structures
 that are intrinsic to $BC_4$.  So these are ``$BC_4$-centric'' 
 definitions of the chiral, vector, and tensor multiplet adinkras 
that do not require any information from the higher dimensional 
supermultiplets.  Only the behavior of the $BC_4$ elements under 
the $\bm *$-map and the holoraumy tensors associated with each 
$BC_4$ element have been used to define the respective adinkras 
to be associated with each off-shell supermultiplet.  Although there
is nothing in these definitions that depend on structures outside
$BC_4$, the requirement on the commutativity or non-commutativity 
of the various holoraumies is motivated by the study \cite{HoLoRmY4D}
where these conditions were found to hold in four dimensional
description of these supermultiplets. 
 
 The arguments are a little bit more involved if one begins the analysis
 from a starting point of picking a element of $BC_4$ that depends on
 odd permutations.  But with appropriate modifications, the same final
 result occurs.
 
 In this chapter, we have proposed a set of criterion and described
 a process by which one-half of all possible four color adinkras 
 described by $BC_4$ can simultaneously describe results obtainable 
 from a 0-brane reduction procedure applied to minimal off-shell 4D, 
 $\cal N$ = 1 supermultiplets.

\section{Conclusion}
\label{conclusions}

Before leaving entirely the realm of conjectures, there is one more
that we would like to present.  This one is not confined to adinkras
related to $BC_4$.

In a recent fascinating development \cite{adnkGEO} in this general line 
of research on Garden Algebras, adinkras, and codes (GAAC), there has 
appeared indications that adinkras can be interpreted as objects possessing
algebraic geometrical descriptions as punctures of Riemann surfaces.   
With this occurrence, we conjecture the holoraumy matrices defined 
by $ {\bm {\Tilde V}_{\rI\rJ}^{(\cal R)}}$ will also likely be related to a 
construction with a basis in algebraic geometry.

$~~~$ {\it {Conjecture \# 3}}:  \newline \indent $~~~$
{\it {The holoraumy matrices}} $ {\bm {\Tilde V}_{\rI\rJ}^{(\cal R)}}$ {\it {that
can be constructed from the L-matrices}} \newline \indent $~~~$
{\it {and R-matrices of the ``Garden Algebra'' may be obtained from
an algebraic}} \newline \indent $~~~$
{\it {geometrical construction based on monodromy matrices. }} \vskip,2in

In this paper, we have attempted to repeat the path pioneered by the
work in \cite{BwT1} that showed how adinkras in one dimension can be 
extended to understand when such adinkras also allow the interpretation 
of being the reductions of 2D, $\cal N$ = 1 supermultiplets.  The work 
in \cite{BwT1} can be interpreted as the analog of the integration of a 
1-cycle along a closed path.  More recently \cite{2adnk}, however, there 
has been introduced another methodology only based on the codes.
Older works, \cite{codes} had made note of the role of codes in defining
irreducible representations of adinkras that descend from four dimensions.
But the work in \cite{2adnk} emphasizes that codes also play a role
in understanding dimensional enhancement from 1D to 2D.  In the
light of the result in \cite{adnkKyeoh} on fermionic dimensional 
enhancement, it would be an interesting investigation to see how
codes play a role in that result.

If the conjectures made in this paper are valid, the path now seems
open for deriving how adrinkras and restriction places there upon
give rise to supersymmetrical representations in {\it {all}} higher
dimensions.

 \vspace{.05in}
 \begin{center}
\parbox{4in}{{\it ``A mathematician, like a painter or a poet, is a maker 
of patterns. If his patterns are more permanent than theirs, it is because 
they are made with ideas.'' \\ ${~}$ 
 \\ ${~}$ 
\\ ${~}$ }\,\,-\,\, G.\ H.\ Hardy $~~~~~~~~~$}
 \parbox{4in}{
 $~~$}  
 \end{center}
 
  \noindent
{\bf Acknowledgements}\\[.1in] \indent
This work was partially supported by the National Science Foundation grant 
PHY-13515155.  Additional acknowledgment is given by D.E.A.G.\ for a Graduate 
Prize Fellowship at Harvard Univ., to the CSPT, as well as recognition for her 
participation in 2013-2015 SSTPRS (Student Summer Theoretical Physics 
Research Session) programs.   D.E.A.G. also thanks the organizers of the 
2015 Miami Topical Physics Conference for the invitation to present.  S.J.G. 
acknowledges the generous support of the Roth Professorship and the very 
congenial and generous hospitality of the Dartmouth College physics department.  
This research was also supported  in part the University of Maryland Center for 
String \& Particle Theory (CSPT).

$$~~$$

\end{document}

\bibliographystyle{elsart-numX}
\small\raggedright
\bibliography{Refs}

\end{document}